\documentclass[11pt]{article}

\usepackage[utf8]{inputenc}
\usepackage{fullpage}
\usepackage{amsmath} 
\usepackage{bm} 
\usepackage{booktabs} 
\usepackage{threeparttable}
\usepackage{subcaption} 
\usepackage{lineno}
\usepackage{graphicx}
\usepackage{natbib}
\usepackage{authblk}
\usepackage{setspace}
\providecommand{\keywords}[1]{\textit{\textbf{Key words}}: #1}

\begin{document}

\title{Bayesian spatial modelling of terrestrial radiation in Switzerland}

\author[1]{Christophe L. Folly}
\author[1,2]{Garyfallos Konstantinoudis}
\author[1]{Antonella Mazzei-Abba}
\author[1]{Christian Kreis}
\author[3]{Benno Bucher}
\author[4]{Reinhard Furrer}
\author[1]{Ben D. Spycher\footnote{Corresponding author; e-mail: ben.spycher@ispm.unibe.ch}}

\affil[1]{Institute for Social and Preventive Medicine (ISPM), University of Bern, Switzerland}
\affil[2]{MRC Centre for Environment and Health, Department of Epidemiology and Biostatistics, School of Public Health, Imperial College London, London, UK}
\affil[3]{Swiss Nuclear Safety Inspectorate, Brugg, Switzerland}
\affil[4]{Department of Mathematics and Department of Computational Science, University of Zurich, Zurich, Switzerland}

\date{} \setcounter{Maxaffil}{0}
\renewcommand\Affilfont{\itshape\small}
\maketitle

\normalsize

\begin{abstract}
 \vspace*{-.5em} The geographic variation of terrestrial radiation can be exploited in epidemiological studies of the health effects of protracted low-dose exposure. Various methods have been applied to derive maps of this variation. We aimed to construct a map of terrestrial radiation for Switzerland. We used airborne $\gamma$-spectrometry measurements to model the ambient dose rates from terrestrial radiation through a Bayesian mixed-effects model and conducted inference using Integrated Nested Laplace Approximation (INLA). We predicted higher levels of ambient dose rates in the alpine regions and Ticino compared with the western and northern parts of Switzerland. We provide a map that can be used for exposure assessment in epidemiological studies and as a baseline map for assessing potential contamination. 
\end{abstract}


\section{Introduction}
Terrestrial radiation stems from radio nuclei contained in the topsoil. The main contribution comes from $^{40}$K and the elements of the Uranium and Thorium decay chains. The levels of ambient dose rates from terrestrial origin vary in space, depending on the local concentrations of the respective radioactive isotopes in the soil \citep{united2018sources}. \\

This geographic variation has been exploited in epidemiological studies on the health effects of protracted low dose exposures. Recent work has specifically looked at possible links between risk of childhood cancer and background ionising radiation \citep{mazzei2020, demoury2017, spix2017, kendall2013}. The effects on cancer risks of protracted low-dose ionising radiation are expected to be small and require large sample sizes to be detected. As direct measurements of doses are not feasible for large numbers of study participants, these studies assessed exposures using geographic exposure models and geocoded residential address information.  \\

A variety of methods have been used to develop geographical models for terrestrial background radiation for exposure assessment. In Great Britain, an ordinary least squares (OLS) regression including predictors such as bedrock classes and radiation measurement means by county provided the best performance \citep{chernyavskiy2016, kendall2016} among various modelling approaches. In France, \citet{warnery2015} applied ordinary kriging and multi-collocated co-kriging to a large dataset of dosimetry measurements conducted in veterinary clinics. The two methods performed similarly. However, the co-kriging approach, which jointly modelled geogenic uranium potential with terrestrial radiation, showed more detailed spatial features. In Switzerland,  \citet{rybach1997, rybach2002} used inverse distance interpolation to derive maps of terrestrial gamma radiation from naturally occurring radionuclides and from $^{137}$Cs (fallout from the Chernobyl nuclear accident) based on a heterogeneous set of measurements including in situ measurements, laboratory measurements of stone samples and airborne $\gamma$-spectrometry. These estimates were combined with dose rates from cosmic radiation, calculated as a function of elevation, to obtain a map of total external background radiation.\\

Integrated nested Laplace approximations (INLA) used together with stochastic partial differential equations (SPDE) allow fitting models involving Gaussian random fields (GRF) to large data sets at reasonable computation costs by establishing an explicit link between the GRF and Gaussian Markov random fields (GMRF). This method has for instance been applied to the spatial prediction of soil pH and elemental concentrations \citep{huang2017} and the spatio-temporal modelling of air pollution \citep{cameletti2013}. \\

Here our goal was to construct an improved map of terrestrial radiation in Switzerland as a tool for exposure assessment in epidemiological studies. Since the work of \citet{rybach1997}, a more extensive set of radiation measurements has become available. Together with advances in spatial statistics and increased computing power, these should allow the development of more accurate maps of terrestrial radiation than those currently available for Switzerland. \\

\section{Data}

\subsection{Radiation measurements}

We used airborne $\gamma$-spectrometry measurements, which were carried out for various purposes including: regular flights to monitor the areas around nuclear facilities (Fig. \ref{fig:flightpath}), survey flights to collect reference values in areas of high population density, and training flights for source detection and for international intercomparison exercises. Furthermore, flights traversing large sections of the country as well as targeted flights to observe local anomalies in background radiation have been performed (Fig. \ref{fig:folds}). \\

The measurements were performed from a helicopter flying at a height of about 90 m above ground. The system consisted of an a 16.8 litres NaI detector and a spectrometer with 256 channels and energy range of 40--3'000 keV. A spectrum was recorded each second. The ambient dose rate 1 m above ground was computed based on the recorded spectra using the spectrum dose index (SDI) method, described in detail in \citet{bucher2001}. The SDI method has been calibrated by dose rate measurements on ground. The field of view of the detector corresponds to a surface of about $300\times300$ m\textsuperscript{2} on ground. This results in a dense coverage and a large overlap of sequential measurements.\\

The measurements have a relative uncertainty of about 10\% for terrestrial radiation. Windows of the spectra allow inferring the soil concentration of $^{40}$K and the elements of the uranium and thorium decay chains, albeit with larger uncertainty. Measurements for $^{137}$Cs mostly lie below the detection limit, however the contribution of $^{137}$Cs to ambient dose rates is incorporated in the estimates for the terrestrial radiation in the SDI method. \\

\subsection{Predictors}
As potential predictors of ambient dose rates, we considered information on the local geology and land coverage. Geological maps of Switzerland were obtained from the Federal Office of Topography (Swisstopo). We included information on the tectonic plate (19 categories) and on the lithology (5 categories) of surface rock \citep{swisstopo2005geolmaps}. Tectonic information captures large scale geological units to which the local geological formations belong. Lithological information refers to the geological processes of rock formation including sedimentary, metamorph or magmatic formation processes as well as categories for unconsolidated rocks and glaciers. Individual categories are listed in tables \ref{tab:coeffs_litho},\ref{tab:coeffs_tecto} and \ref{tab:coeffs_lc}.\\

Land coverage information was extracted from the areal statistics obtained from the Federal Statistical Office (FSO). Grid cells of 100$\times$100 m$^{2}$ are classified into six pre-defined classes based on areal photographs taken during the period of 2004--2008. The six classes refer to artificial surface covers, three vegetation categories (grass, bushes, trees), natural surfaces without vegetation cover and water and wetland surfaces.\\

We included daily rainfall data \citep{testbedmeteoswiss} after the Chernobyl accident as we still expect contributions to the ambient dose rate from $^{137}$Cs contamination from the aftermath of the accident. Most highly affected areas lie in the canton of Ticino. We aggregated the rainfall over the days from 30 April until 5 May 1986. The choice of the days is based on air filter measurements in Fribourg (western Switzerland) and on an animation of the atmospheric spread of the radioactive cloud by the IRSN \citep{albergel1988radioactiveplume}.\\

\section{Methods}
\subsection{Data cleaning and thinning out}
\begin{figure}[t]
    \centering
    \includegraphics[width=0.7\linewidth]{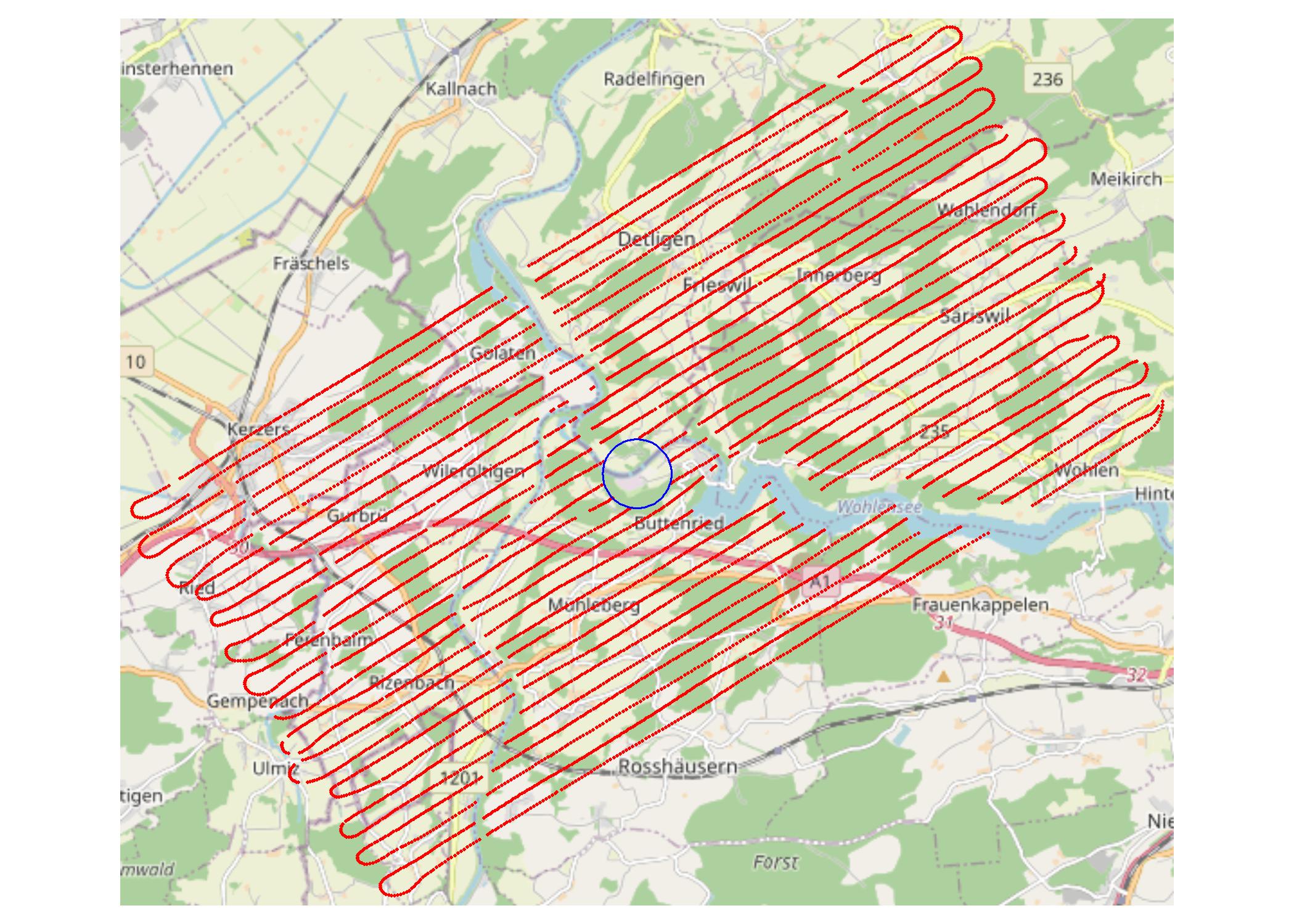}
    \caption{Location of measurements (red dots) made during a measurement flight over the nuclear power plant in Mühleberg. The shown points include all measurements (without thinning) after removing those made in the vicinity the nuclear facility (circle in the center) as well as those made over water bodies.}
    \label{fig:flightpath}
\end{figure}

We removed values influenced by artificial point sources and measurements at locations where the terrestrial radiation was shielded by water bodies (Fig. \ref{fig:flightpath}). Point sources include nuclear power plants, the research site of the Paul-Scherrer Institute (including the intermediate storage facilities for nuclear waste), and a building belonging to the former test reactor facility in Lucens. We investigated spatial correlation with variograms. \\ 

Consecutive measurements are correlated due to the overlap of the field of view. This overlap introduces an additional correlation structure on the observation level and masks the inherent correlation of the underlying spatial process \citep{2013arXiv1302.4659H}. We thinned out the measurements by considering only every fifteenth measurement. Details with regard to the thinning out and the reasoning behind it are provided in \ref{App:thinning_out}. \\

\subsection{Statistical model}

\subsubsection{Model definition}
We modelled the ambient dose rate using the following log-linear mixed-effects model \begin{equation}\label{eq:mod_std}
    Y(\bm{s}) = \bm{X}(\bm{s})\bm{\beta}  + U(\bm{s}) + \epsilon(\bm{s}) 
\end{equation}
where $Y(\bm{s})$ is the log-transformed dose rate at location $\bm{s}$, $\bm{\beta}$ is a vector containing the coefficients of the fixed effects of covariates $\bm{X}(\bm{s})$, $U(\bm{s})$ is a Gaussian random field (GRF) with Mat\'{e}rn covariance function and $\epsilon(\bm{s})$ is white noise with variance $\sigma_{\epsilon}^{2}$. \\

The Mat\'ern covariance function has a scale parameter $\kappa > 0$ and smoothness parameter $\nu > 0$. It is defined as \begin{equation}
    Cov(U(\bm{s}_{i}), U(\bm{s}_{j})) = \sigma_{U}^{2}\frac{2^{1-\nu}}{\Gamma(\nu)}(\kappa \parallel \bm{s}_{i} - \bm{s}_{j} \parallel)^{\nu}K_{\nu}(\kappa \parallel \bm{s}_{i} - \bm{s}_{j} \parallel)
\end{equation}
where $\parallel \bm{s}_{i} - \bm{s}_{j} \parallel$ is the Euclidean distance between locations $\bm{s}_{i}$ and $\bm{s}_{j}$, $K_{\nu}$ the modified Bessel function of the second kind, and $\sigma_{U}^{2}$ is the marginal variance of the spatial process $U(\bm{s})$.\\

Formulated as a hierarchical Bayesian model, we write: \begin{equation}\label{eq:mod_hierar}
    \begin{split}
    Y(\bm{s})|\bm{\beta},U(\bm{s}),\sigma_{\epsilon} & \sim N(\bm{X}(\bm{s})\bm{\beta}  + U(\bm{s}), \sigma_{\epsilon}^{2}) \\
    U(\bm{s})|\bm{\Theta} & \sim \bm{GRF}(\bm{0},\bm{\Sigma}) \\
    \bm{\Theta},\bm{\beta},\sigma_{\epsilon} & \sim \Pi 
    \end{split}
\end{equation}
where $\Theta$ is a vector of parameters of the GRF and $\bm{\Sigma}$ its covariance matrix with $\Sigma_{ij} = \sigma_{U}^{2}Cor(U(\bm{s}_{i}), U(\bm{s}_{j}))$ and $\Pi$ denotes the prior distributions.  \\

\subsubsection{Inference}
Fitting models involving GRFs becomes cumbersome with increasing number of measurements, as the computation involves inverting large matrices \citep{lasinio2013}. As work-arounds, various approximations have been proposed both within the frequentist and Bayesian framework. These include, for example,  covariance tapering \citep{furrer2006} fixed-rank kriging \citep{cressie2008}, and an approach based on SPDEs \citep{lindgren2011}. A comparison of proposed techniques suggested similar performance of these methods in a case study competition\citet{heaton2019}.  \\

We applied the SPDE approach \citep{lindgren2011} and fitted the model using integrated nested Laplace approximations \citep{rue2009}. Integrated nested Laplace approximations (INLA) is a deterministic alternative to Markov chain Monte Carlo (MCMC) methods for fitting (latent Gaussian) Bayesian models\citep{rue2009}. In the SPDE approach, the Gaussian random field is linked explicitly to a Gaussian Markov random field through the solution $U(\mathbf{s})$ of the SPDE
\begin{equation}\label{eq:spde}
    (\kappa - \Delta)^{\alpha/2}U(\mathbf{s}) = W(\mathbf{s}),\: \text{with} \;   \mathbf{s} \in R^{d},\; \alpha = \nu+d/2,\; \kappa > 0,\; \nu > 0
\end{equation}
where $W$ is Gaussian white noise, $\Delta$ is the Laplace operator, and $U(\bm{s})$ is a continuous GRF with Mat\'ern covariance \citep{whitle1954stationary}. \citet{lindgren2011} represent a weak solution to eq. (\ref{eq:spde}) as a Gaussian Markov random field (GMRF) using the finite element method, expressing the GMRF as linear combination of basis functions defined on the nodes of a triangular mesh. This allows for an explicit link between the GRF and GMRF. A more detailed description can be found in \citep{krainski2018advanced}. Exact solutions are obtained on the nodes of the mesh and linearly interpolated to a continuous field. \\

To fit our models, we used the dedicated package R-INLA in the R computing environment  \citep{martins2013,lindgren2015}(www.r-inla.org). In R-INLA, the smoothness parameter $\nu$ of the Mat\'ern covariance function is coupled to the parameter choice in the SPDE, and does not need to be additionally set. We chose $\alpha = 2$, corresponding to $\nu = 1$ (since $d=2$). The R-INLA package provides built-in functions to construct the mesh. More nodes, which result in a denser mesh, allow for a smoother representation of the field, but increase the computational burden. The mesh density was tuned by providing the minimal (3.5km) and maximal distances (5km) between nodes together with the minimal angle (31$^{\circ}$) between edges. \\

To define priors for the hyper parameters of the spatial field $U(\bm{s})$ ($\sigma_{U}$ and $\kappa = \sqrt{8\nu}/\rho$, where $\rho$ is the (practical) range of the spatial correlation), we used penalized complexity priors (PCpriors) \citep{simpson2017}. These priors penalize the complexity of the model by shrinking the standard deviation of the field to zero and shrinking the range of the Gaussian field towards infinity. We specified the priors through $P(\sigma_{U} > 10)= 0.01$ for the standard deviation of the field, assuming that it is unlikely to have such high spatial variation of radiation left after adjusting for the covariates, and $P(\rho > 15) = 0.5$ for the range of the field, expecting the range to be in the order of 10--20km after exploring the variograms. Normal priors with mean zero and precision 0.001 were used for the fixed-effects $\bm{\beta}$ and an inverse gamma prior with shape 1 and scale $5\times10^{5}$ for the precision ($1/\sigma^{2}_{\epsilon}$). \\

\subsection{Extended model}
To allow for a more complex spatial correlation structure, we extended the model by adding a second spatial random effect $U_{2}(\bm{s})$. The extended model can be written as \begin{equation}\label{eq:extend}
    Y(\bm{s}) = \bm{X}(\bm{s})\bm{\beta} + U_{1}(\bm{s}) + U_{2}(\bm{s}) + \epsilon(\bm{s})
\end{equation}
where we intended $U_{1}$ and $U_{2}$ to capture both large scale correlation allowing to make predictions for areas not covered by in areal survey and short range variation helping to fit the data better in densely surveyed areas. \\

To fit the model, we chose the priors for the field $U_{1}$ as $P(\sigma_{U_{1}} > 10)= 0.01$ and $P(\rho_{U_{1}} > 15) = 0.6$, the priors for $U_{2}$ as $P(\sigma_{U_{2}} > 10)= 0.01$ and $P(\rho_{U_{2}} > 2) = 0.02$, forcing $U_{2}$ towards short range correlations. 

\subsection{Model comparison}
We compared the mixed-effects model and the extended mixed-effects model to simpler models. We were interested in whether the added complexity of the (extended) mixed-effects model results in a better predictive performance than the simpler model \ref{eq:mod_std}. We also considered the following simplified models:\begin{itemize}
    \item Bayesian linear regression (fitted with INLA): $$ Y(\bm{s}) = \bm{X}(\bm{s})\bm{\beta} + \epsilon(\bm{s})$$ 
    \item Spatial Bayesian random-effects model: $$ Y(\bm{s}) = U(\bm{s}) + \epsilon(\bm{s}) $$ 
\end{itemize}
with $Y(\bm{s})$, $\bm{\beta X}$, $\epsilon$ and $U(\bm{s})$ as above. The same set of covariates, parameters for mesh construction, and priors were chosen for the simplified models.\\

To assess model performance we conducted two types of cross-validation: 1) Random cross-validation by randomly splitting the data into a training (70 percent) and a validation set (30 percent), and 2) Spatial cross-validation by partitioning the country into grid cells of $15\times15$ km, assigning each cell randomly to one out of four folds and recomputing the model four times, each time leaving out one of the folds. Assignment to the folds was done using the R package blockCV \citep{valavi2018blockcv}. \\

The spatial cross-validation was tailored to a typical range of extrapolation from surveyed into non-surveyed areas that would be required for exposure assessment in epidemiological studies. The size of the spatial blocks was chosen based on the distance of residential address geocodes to the closest measurement. Geocoded locations of residential buildings in Switzerland were obtained through the Swiss National Cohort \citep{bopp2009cohort}. We considered locations further than 250 m away from the closest measurement and chose the side length of a block as twice the 90 percent quantile of the distribution of distances between residential locations and closest measurement. \\

As performance measure for comparing models, we considered the root mean square error defined as \begin{equation}\label{eq:rmse}
    \text{RMSE} = \left[\frac{1}{N} \sum_{i = 1}^{N}(y_{p_{i}} - y_{m_{i}})^{2}\right]^{1/2}
\end{equation}
where the indices $p_i$ and $m_i$ stand for predicted and measured values at the locations $i=1,\dots,N$ for which predictions are made. We considered the mean of the predicted posterior $y_{p}$ as point prediction in order to facilitate the interpretation. The RMSE and the $R^{2}$ (defined as $Cor(Y_{p},Y_{m})^{2}$) are separately reported for the model fit to both the training and validation folds.\\

In addition, we computed the continuous ranked probability score (CRPS) \citep{gneiting2007}, a measure of predictive accuracy defined as \begin{equation}\label{eq:crps}
    \text{CRPS}(F,y) = \frac{1}{N} \sum_{i = 1}^{N} \int_{-\infty}^{\infty}\left[F_i(x) - 1(x \geq y_{m_{i}})\right]^{2}dx
\end{equation}
where $F_{i}()$ is the posterior cumulative distribution function of the prediction at location $i$ and $y_{m_{i}}$ the corresponding measurement. Smaller CRPS are preferred. It penalizes both deviation of the predicted from the measured value as well as large predictive uncertainty. To facilitate computation we assumed normality of the posterior distributions. We extracted means and standard deviations of the pointwise posteriors obtained from INLA and calculated the CRPS using the R package scoringRules \citep{jordan2017}. \\

\subsection{Sensitivity analysis}
For the selected model, we assessed whether modifications of the proportion  of measurements kept during thinning out (every $10\textsuperscript{th}$ and every $20\textsuperscript{th}$ vs. every $15\textsuperscript{th}$ in the main model) and of the mesh (denser and less dense than main model) affected the results. We compared the resulting predictions on a $1\times1$ km grid by computing the $R^{2}$ between the means of pointwise posteriors, and examined changes in the resulting hyper parameters and coefficients. 

\subsection{Predictions}
We computed posterior mean, mode, and standard deviation of terrestrial radiation on a $100\times100$ m\textsuperscript{2} square grid over Switzerland using the best performing model. \\

Calculations were performed on UBELIX (http://www.id.unibe.ch/hpc), the HPC cluster at the University of Bern, and the CX1 Imperial College London cluster.

\section{Results}

\subsection{Description of measurements}
\begin{figure}[t]
    \centering
    \includegraphics[width=0.7\linewidth]{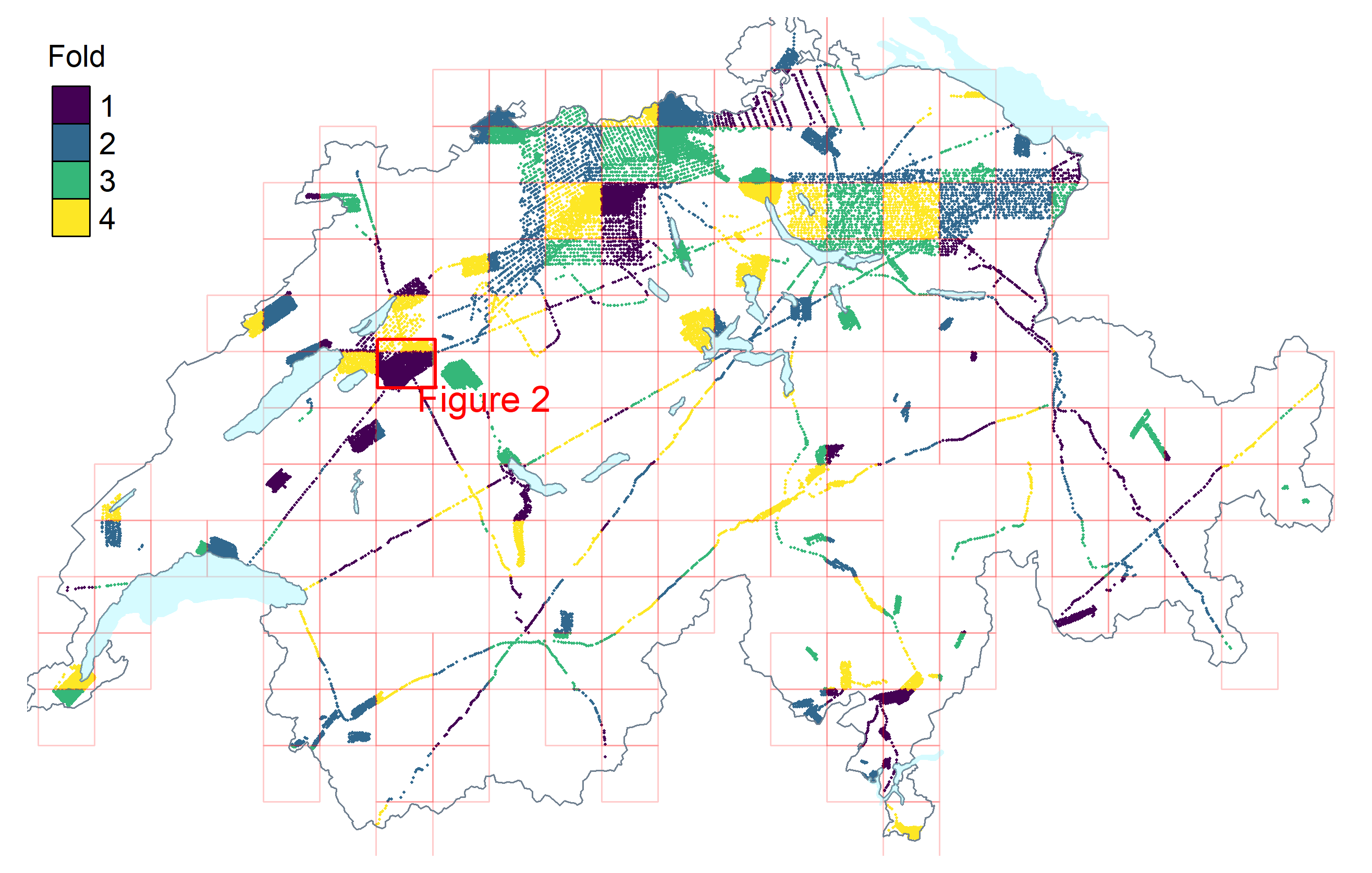}
    \caption{Coverage and partition into four folds of the airborne radiation measurements used for model fitting. Displayed are the 39'258 measurements retained after data thinning (every 15th measurement). These measurements were made during 214 flights. The red box indicates the area displayed in Fig. \ref{fig:flightpath}.}
    \label{fig:folds}
\end{figure}

The database consisted of 654'530 measurements made during 214 flights. After cleaning, inculding removal of those influenced by point sources or water bodies, 589'197 measurements were available. The measurements densely cover large parts of the northern part of Switzerland, where population density is highest. The coverage and the partition into 4 spatial folds are illustrated in Figure \ref{fig:folds}. \\

The measured ambient dose rates from terrestrial radiation ranged from 7.02 nSv/h to 417.13 nSv/h (mean: 54.25 nSv/h,  median: 50.94 nSv/h, interquartile range [IQR] of 42.12--61.74 nSv/h). After thinning out (keeping only every $15\textsuperscript{th}$ measurement), the range was 7.60 to 398.38 nSv/h (mean: 54.24 nSv/h, median: 50.87 nSv/h, IQR: 42.12--61.63 nSv/h). The highest dose rates were measured in mountainous areas with bare rock surfaces. Low dose rates were measured over water bodies and, particularly, above glaciers, where the ice has a shielding effect. \\

\begin{figure}[t]
    \centering
    \includegraphics[width=0.6\linewidth]{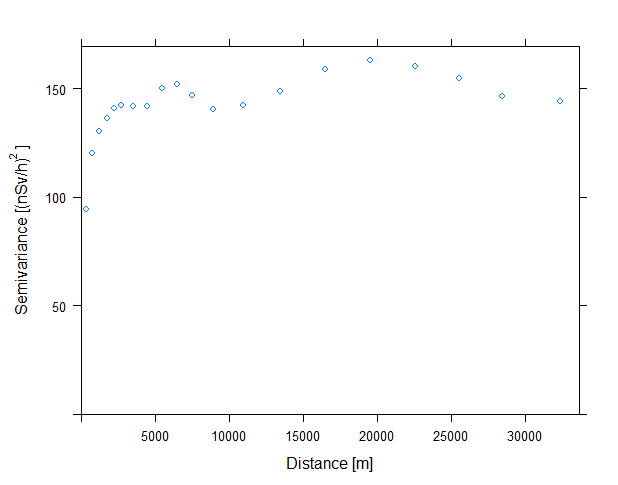}
    \caption{Semivariogram for the ambient dose rates from terrestrial radiation after inclusion of an external trend. The variogram was computed based on 100'000 measurements randomly sampled from the cleaned measurement data base.}
    \label{fig:vrg}
\end{figure}

Variograms before and after including an external trend revealed no (global) directional anisotropy, but indicated spatial processes at different ranges (Fig. \ref{fig:vrg}), motivating the extension of the mixed-effects model with a second spatially correlated GRF. \\

\subsection{Comparison of fitted models}
The fitted hyper parameters of the different models are shown in Table \ref{tab:res_hyper}. The practical range was estimated to be 13.9 km in the mixed-effects model, and 15.3 km in the pure spatial model. In the extended model, the fitted practical ranges were 26.6 km for $U_{1}$ and 1.7 km for $U_{2}$. The resulting $\beta$'s are displayed in Tables \ref{tab:coeffs_litho}, \ref{tab:coeffs_tecto}, \ref{tab:coeffs_lc}, and \ref{tab:coeffs_rain} (Appendix). \\

\begin{table}[!b]
\centering
\begin{threeparttable}
\caption{Mean and standard deviation of marginal posteriors of the hyper parameter for a linear model, a purely spatial random-effect model, the standard mixed-effects model and the extended mixed-effected model including a second spatially correlated random-effect. }
\begin{tabular}[t]{lrrr}
\toprule
Model & (Hyper) Parameter & Mean & Standard Deviation \\
\midrule
Linear Model & Precision $1/\sigma_{\epsilon}^{2}$ & 12.85 & 0.09 \\
Random effects & Precision $1/\sigma_{\epsilon}^{2}$ & 19.32 & 0.15\\
& Range\tnote{*}\hspace{0.15cm} $U$ [km] & 15.27 & 1.63\\
& $\sigma_{U}$ & 0.434 & 0.024\\
Mixed-effects & Precision $1/\sigma_{\epsilon}^{2}$ & 20.94 & 0.15\\
& Range\tnote{*}\hspace{0.15cm} $U$ [km] & 13.9 & 1.2 \\
& $\sigma_{U}$ & 0.383 & 0.019\\
Extended mixed-effects & Precision $1/\sigma_{\epsilon}^{2}$ & 36.04 & 0.31 \\
& Range\tnote{*}\hspace{0.15cm} $U_{1}$ [km] & 26.64 & 4.02 \\
& $\sigma_{U_{1}}$ & 0.322 & 0.025 \\
& Range\tnote{*}\hspace{0.15cm} $U_{2}$ [km] & 1.69 & 0.06 \\
& $\sigma_{U_{2}}$ & 0.181 & 0.003 \\
\bottomrule
\end{tabular}
\label{tab:res_hyper}
\begin{tablenotes}
        \item[*]{Practical range, parametrized as $\rho = \sqrt{8\nu}/\kappa$, where $\kappa$ is the scale parameter and $\nu$ the smoothness parameter (fixed as $\nu = 1$) of the Mat\'ern covariance function.}
    \end{tablenotes}
    \end{threeparttable}
\end{table}

The extended mixed-effects model performed best in both cross-validation settings and across all measures (Table \ref{tab:performance}). It achieved an $R^{2}$ of 0.4 and CRPS of 0.153 in the spatial cross-validation compared to an $R^{2}$ of 0.27 and CRPS of 0.217 of the standard mixed-effects model. As expected, all models performed better in a random cross-validation than in a spatial cross-validation setting, where measurements of the validation and training sets are further apart. The differences in performance between the validation settings were more pronounced for the models including spatially correlated random-effects than for the linear model. Moreover, while the linear model is clearly outperformed by the other models in the random cross-validation, only the extended mixed-effects model performs considerably  better than the simple linear model in the spatial cross-validation setting.\\

\begin{table}[!b]
    \centering
    \begin{threeparttable}
    \caption{Measures of predictive performance of the point-wise posterior mean terrestrial radiation from different models in the random\tnote{a} and spatial\tnote{b} cross-validation.}
    \label{tab:performance}
    \begin{tabular}{lllllll}
        \toprule
             &  \multicolumn{3}{c}{random CV\tnote{a}}& \multicolumn{3}{c}{spatial CV\tnote{b}}\\
            Model & RMSE & $R^{2}$ & CRPS & RMSE & $R^{2}$ & CRPS \\
        \midrule
        Linear model & 15.89 & 0.39 & 0.205 & 17.41 & 0.32 & 0.220\\
        Random-effect & 12.53 & 0.62 & 0.162 & 19.89 & 0.22 & 0.236\\
        Mixed-effects & 12.02 & 0.64 & 0.154 & 18.85 & 0.27 & 0.217\\
        Extended mixed-effects & 10.17 & 0.75& 0.115 & 16.53 & 0.40 & 0.153\\
        \bottomrule
    \end{tabular}
    \begin{tablenotes}
        \item[*]{CV, Cross-validation; CRPS, continous ranked probability score; RMSE, root mean square error}
        \item[a]{The data was split randomly into a training (70 percent) and validation (30 percent) sample.}
        \item[b]{The country was partitioned into 15$\times$15km blocks, which then were assigned randomly to one out of four folds (Fig. \ref{fig:folds}). The model was recomputed four times, each time leaving out one of the four folds for validation. Displayed are the performance measures averaged over the four folds.}
    \end{tablenotes}
    \end{threeparttable}
\end{table}

\subsection{Sensitivity analysis}
When fitting the extended mixed-effects model to different subsets of the data (every 10th, and every 20th measurement included), the resulting predictions did not change substantially (Table \ref{tab:sens_hyper}). The $R^{2}$ between the predictions based on these subsets and the subset used in the main analysis were 0.96 and 0.97 respectively. When fitting the model using a less dense or a denser mesh, we obtained an $R^{2}$ larger than 0.99. The fitted hyper parameters are similar (Table \ref{tab:sens_hyper}). 

\begin{table}
    \centering
    \begin{threeparttable}[!b]
    \caption{The fitted hyper parameters (posterior mean) in sensitivity analysis with regard to the thinning out of the measurements and with regard to the mesh density compared the hyper parameters fitted in the main analysis.}
    \begin{tabular}{llllll}
    \toprule
    Model & original\tnote{a} & thin10\tnote{b} & thin20\tnote{c} &  denser mesh\tnote{d} & less dense mesh\tnote{e} \\
    \midrule
    Precision $1/\sigma_{\epsilon}^{2}$ & 36.0 & 36.5 & 35.2 & 32.6 & 37.5\\
    Range $U_{1}$ [km] & 26.6 & 22.5 & 27.4 & 28.2 & 23.5\\
    Standard deviation $U_{1}$ & 0.32 & 0.37 & 0.32 & 0.30 & 0.34\\
    Range $U_{2}$ [km] & 1.69 & 1.38 & 1.42 & 1.96 & 1.52 \\
    Standard deviation $U_{2}$ & 0.18 & 0.20 & 0.19 & 0.18 & 0.18\\
    \bottomrule
    \end{tabular}
    \label{tab:sens_hyper}
    \begin{tablenotes}
     \item[a]{We used every fifteenth measurement and meshs with 4'928 and 202'357 nodes for $U_{1}$, resp. $U_{2}$, to fit the model and compute predictions.}
     \item[b]{Instead of every fifteenth, we used every tenth measurement.}
     \item[c]{Instead of every fifteenth, we used every twentieth measurement.}
     \item[d]{Meshs with 10'880 and 255'916 nodes were used.}
     \item[e]{Meshs with 2'327 and 114'517 nodes were used.}
    \end{tablenotes}
    \end{threeparttable}
\end{table}

\subsection{Predicted dose rates}
\begin{figure}[!t]
\includegraphics[width=0.9\linewidth]{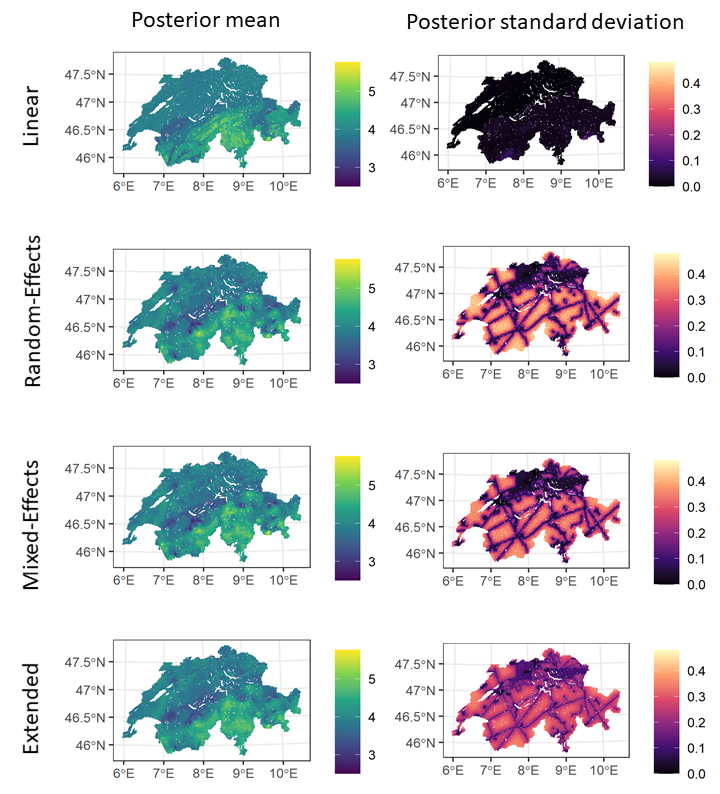} 
\caption{Maps of pointwise posterior mean (predicted mean, left) and standard error of predictions (right) for log-transformed dose rates from terrestrial gamma radiation based on the linear regression, random-effects, mixed-effects and the extended mixed-effects models. Note that the standard error of the predictions refers to the predicted mean surface and does not include the Gaussian white noise $\epsilon(\bm{s})$.}
\label{fig:maps}
\end{figure}

Figure \ref{fig:maps} shows the resulting maps of predicted dose rates (pointwise posterior means) and their uncertainty (pointwise posterior standard deviations) from the four fitted models. Predictions from the linear model reflect the patterns of the available covariates. All models showed higher levels in the southern mountain areas and in the Canton of Ticino and lower levels in the Central Plateau, where the geology is dominated by the sedimentary Molasse basin. \\

Tectonic units that are related to higher dose rates are Tertiaere Intrusiva und Extrusiva (0.43(0.13)), Unterostalpine Decken (0.26(0.08)) and Oberostalpine Decken (0.32(0.10)). Surfaces covered by bush (-0.019(0.006)) or tree (-0.038(0.004)) vegetation and wetland (-0.13(0.02)) surfaces are associated with lower dose rates, whereas cumulative rainfall after Chernobyl shows a positive association (5.1(1.8)). The contrast between coefficients for the categorical covariates is shrunk in the mixed-effects and extended mixed-effects model compared to the linear model.\\

\begin{figure}[!t]
    \centering
    \includegraphics[width=0.7\linewidth]{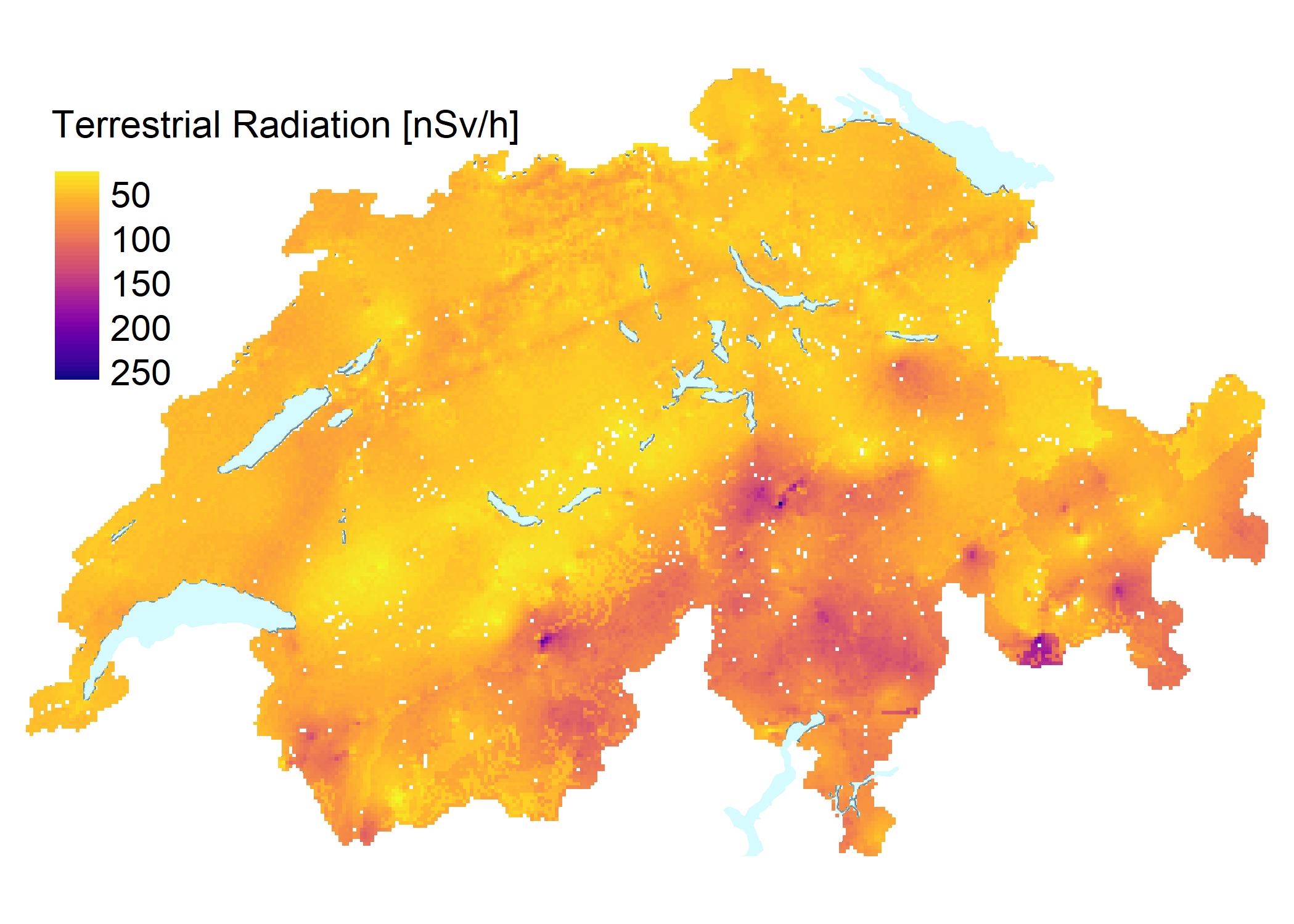}
    \caption{Backtransformed ambient dose rates from terrestrial radiation predicted by the extended mixed-effects model.}
    \label{fig:map_terrRad}
\end{figure}

\begin{figure}[!t]
    \centering
    \includegraphics[width=0.6\linewidth]{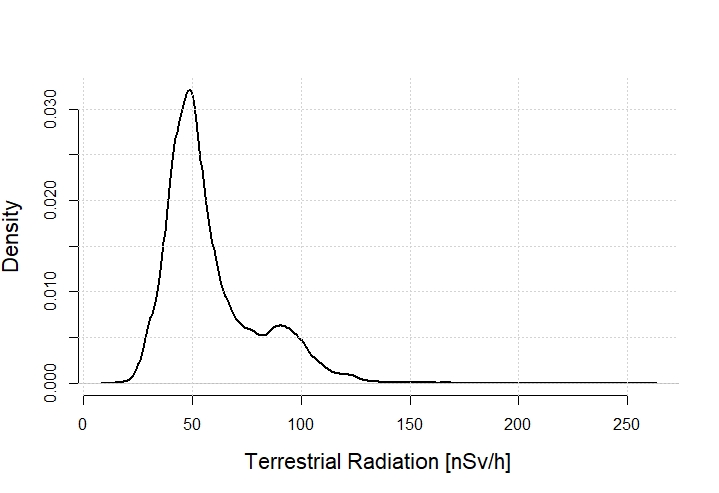}
    \caption{Geographically weighted distribution of dose rates from terrestrial gamma radiation predicted by the extended mixed-effects model.}
    \label{fig:extend_distr}
\end{figure}

Predictions of the preferred extended mixed-effects model, shown in Figure \ref{fig:map_terrRad}, have lower uncertainty in areas not covered by areal survey compared to the simple mixed-effects model (Fig. \ref{fig:maps}, right panel). Due to the coverage of the biggest cities by measurements, uncertainty in the most densely populated areas tends to be lower than for rural areas. The geographically weighted distribution of predicted values of the preferred model shows a narrow concentration around the value of 50 nSv/h with a heavy tail extending to values over 100 nSv/h (Fig. \ref{fig:extend_distr}).\\

\section{Discussion}

We provide new estimates of terrestrial gamma radiation for Switzerland based on an spectrometric areal survey of large parts of the territory and information on land coverage, surface rock and underlying geology. The best performing model includes two spatially correlated random effects capturing short and longer range correlations, respectively.  This model consistently outperformed a model with only one spatially correlated random effect and the simple linear model, particularly in spatial cross-validation, where training and validation sets were spatially separated. \\

Areas whose local geology is dominated by sedimentary or unconsolidated rocks generally exhibit lower dose rates than areas with crystalline rocks. The general pattern displayed by all models is similar to the map of terrestrial radiation developed by \citet{rybach1997}. Contrary to their work, we did not consider the contribution of $^{137}$Cs separately, in view of the levels of radiation stemming from Caesium measured over the last two decades, while the map developed by \citet{rybach1997} reflects levels measured in 1989 and 1990, i.e. relatively shortly after the Chernobyl accident.\\

Previous studies in France and Great Britain applied variants of kriging to model the geographic variation of terrestrial radiation. Conceptually our approach is similar, but it is embedded in a Bayesian framework. While we worked with outdoor measurements, both aforementioned studies focused on indoor measurements. Indoor measurements are influenced by building materials and shielding. \\

Both \citet{warnery2015} and \citet{chernyavskiy2016} reported mean square errors (MSE) from a random cross-validation setting of between 360 to 400 (nSv/h)$^{2}$, which is higher than in our work. \citet{kendall2016} and \citet{chernyavskiy2016} reported the $R^{2}$ in Great Britain to be 0.2--0.27 for different modelling approaches, which is lower than in this work. \\

Measured dose rates had a similar range but a lower mean in Switzerland (mean: 54.25 nSv/h, range: 7--417 nSv/h) compared to France (mean: 76 nSv/h, range: 13--349 nSv/h) \citep{warnery2015}, where these were based on dosimeter measurements from veterinary clinics. Consequently, the range of the predicted values is very similar between the two countries. The fitted range of the spatial correlation is roughly one order of magnitude smaller in Switzerland compared to France. An explanation could be the higher measurement density in our study, allowing us to observe spatial correlations at smaller distances. \\

A strength of our study is the high resolution of measurements in areas covered by areal surveys. However, coverage was patchy leaving large non-surveyed gaps. Single measurements observe a field of view of $300\times300$ m$^{2}$ and thus measured values need to be interpreted as a weighted averaging of the terrestrial radiation levels over the scale of a few 100 square meters. Variations occurring at a smaller scale cannot be captured.\\

We did not separately model the contributions stemming from Caesium-137 as most of the measurements were below the detection limit of the measurement device. Decay and vertical migration of $^{137}$Cs since the Chernobyl accident significantly influenced the observed levels of ambient dose rates, most strongly in Ticino. As the measurements were conducted between 1996 and 2018, the  coefficient for rainfall is expected to capture an average effect from $^{137}$Cs. As most of the measurements have been performed after 2000 and contamination has been relatively low in the rest of Switzerland, we expected the spatial distribution to be stable enough to neglect the temporal trend without loosing relevant features and predictive performance. \\

Several aspects should be considered when using the presented maps for exposure assessment in epidemiological studies. Studies for which exposure over the past few decades is of interest should attempt to separately model the temporal trend of the $^{137}$Cs contribution. Furthermore, it must be kept in mind that our map represents ambient doses outdoors. While this can serve as a proxy measurement for indoor exposure, further research on the differences between indoor and outdoor exposure levels in Switzerland might allow improved estimation of indoor exposure. Due to the gaps in the data coverage, there are areas of relatively large uncertainty with regard to the predicted dose rates. In addition to the map, the chosen modelling approach also allows producing maps of the predictive uncertainty. These could be used to propagate the uncertainty forward to models linking exposure with disease outcomes so as to be correctly reflected in estimates of potential health effects. Besides the exposure assessment in epidemiological studies, the provided map can serve as a baseline map for assessing potential contamination.\\ 

Our finding that two spatial components with differing ranges improved prediction suggests that spatial variation of terrestrial radiation occurs at different scales. The spatial correlation of terrestrial $\gamma$-radiation might thus be better captured by multiple processes acting over different ranges than by a single spatial process. Models that are able to incorporate such behaviour might prove most suitable to map the variations of terrestrial radiation. This also offers the possibility of improving models in future. Similar situations may exist for other exposures. \\

The map, formatted as shapefile, and the R code used for estimating the model are available on Github (https://github.com/FollyCh/TerrMapCH). Also we provide 100 realisations drawn from the joint posterior distribution that can be used for error propagation.

\section*{Acknowledgement}
We thank Dr. Philipp Steinmann from the Federal Office for Public Health for providing information about existing efforts to measure environmental radiation in Switzerland and selecting appropriate data sources. We thank Prof. Moritz Bigalke (Geography, University of Bern) and Dr. Stephan Dall'Angelo from swisstopo for providing information on available information on geological data and assisting our selection of covariates.

\section*{Declaration of competing interest}
The authors have no conflicts of interest to declare.

\section*{Funding}
This work was funded by the Swiss National Science Foundation (Grant No. 320030\_176218) and the Swiss Cancer League (Grant No. KLS-4592-08-2018 ). G.K. is supported by an MRC Skills Development Fellowship (MR/T025352/1).

\newpage
\bibliographystyle{chicago} 
\bibliography{TerrRad_bib}

\appendix
\section{Thinning out}\label{App:thinning_out}
Our decision to thin out the measurements used for model fitting were based on the following considerations.
\begin{enumerate}
    \item \textit{Overlapping observation windows:} consecutive measurements cover partly overlapping areas. Additionally, the measurement errors might be correlated between subsequent measurements, which would introduce spurious correlations. 
    \item \textit{Extreme extrapolations:} when we fitted the models to the full data set, we observed extreme values for extrapolations into unobserved areas, most pronounced around measured local peaks. We assume the reason for these extremes to be trends at the borders of the surveyed areas propagated some distance into the areas not surveyed.
    \item \textit{Redundant information:} As the measurements densely cover surveyed areas, subsets of the data might still contain all relevant information. By thinning out we can save computation time and disk space. Aggregating or thinning out should lead to more stable computations and potential impacts of spurious correlations are mitigated.
\end{enumerate}

Our decision to select only every 15th measurement was based on variograms of consecutive measurements in the directions of flight paths. To compute these, we defined the x-coordinate as the enumeration of subsequent measurement points and the y-coordinate as zero and calculated separate variograms for randomly chosen flights. Results indicated a strong serial correlation of measurements along flight paths which approaches zero only after a lag of 15 to 20 subsequent measurements.

\section{Fitted coefficients} 
\begin{table}[ht]
    \centering
        \begin{threeparttable}
    \caption{Fitted coefficients in the different modelling approaches for categories of the lithology\tnote{*}.}
    \begin{tabular}{llllll}
    \toprule
        & lm inla & mixed & extended \\
        & mean(sd) & mean(sd) & mean(sd)\\
    \midrule
        $\beta_\text{Glacier, Firn}$ & 4.45(0.05) & 3.76(0.06) & 3.74(0.07)  \\
        $\beta_\text{Lockergesteine}$ & 4.21(0.01) & 3.89(0.05) & 3.97(0.05)\\
        $\beta_\text{Magmatische Gesteine}$ & 4.86(0.02) & 4.00(0.05) & 4.01(0.06)\\
        $\beta_\text{Metamorphe Gesteine}$ & 4.36(0.01) & 3.94(0.05) & 3.99(0.05)\\
        $\beta_\text{Sedimentgesteine}$ & 4.21(0.01) & 3.89(0.05) &  3.97(0.05)\\
    \bottomrule
    \end{tabular}
    \label{tab:coeffs_litho}    
    \begin{tablenotes}
     \item[*]{No reference category, as the levels of lithology act as independent intercepts.}
    \end{tablenotes}
    \end{threeparttable}
\end{table}        
        
\begin{table}[ht]
    \centering
    \caption{Fitted coefficients in the different modelling approaches for categories of the tectonic units.}
    \resizebox{\columnwidth}{!}{%
    \begin{tabular}{lllll}
    \toprule
        & lm inla & mixed & extended \\
        & mean(sd) & mean(sd) & mean(sd)\\
    \midrule        
        $\beta_\text{Allochthone Massive und infrapenninische Kristallindecken}$ & reference & reference & reference\\
        $\beta_\text{Ausseralpine Plattform}$  & -0.40(0.01) & 0.06(0.03)& -0.04(0.04)\\
        $\beta_\text{Autochthon - Parauochthon, Infrahelvetische Decken}$ & -0.06(0.02) & 0.02(0.03) & -0.03(0.04)\\
        $\beta_\text{Decken der unterostalpin-penninischen Grenzzone}$ &  -0.30(0.07) & 0.36(0.08) & 0.11(0.07)\\
        $\beta_\text{Faltenjura}$ & -0.42(0.01) & -0.02(0.03) & -0.07(0.03)\\
        $\beta_\text{Helvetische Sedimentdecken s.str.}$ & -0.45(0.02) & -0.10(0.04)& -0.13(0.04)\\
        $\beta_\text{Mittelpenninische Kristallindecken}$ & -0.05(0.02) & 0.11(0.04) & 0.08(0.05)\\
        $\beta_\text{Mittelpenninische Sedimentdecken und -schuppen}$ & -0.75(0.03) & -0.07(0.05)& -0.16(0.05)\\
        $\beta_\text{Molassebecken}$ & -0.39(0.01) & 0.03(0.03) & -0.04(0.03) \\
        $\beta_\text{Oberostalpine Decken}$ & -0.04(0.02) & 0.27(0.09) & 0.32(0.10)\\
        $\beta_\text{Oberpenninische Sedimentdecken}$ & -0.63(0.03) & -0.01(0.05) & -0.06(0.05)\\
        $\beta_\text{Ophiolithfuerende oberpenninische Sedimentdecken und - schuppen}$ & -0.71(0.03) & -0.07(0.05) & -0.15(0.06) \\
        $\beta_\text{Quartaer}$ & -0.35(0.01) & -0.01(0.03) & -0.06(0.03)\\
        $\beta_\text{Sued- bis ultrahelvetische Sedimentdecken und -schuppen}$ & -0.38(0.03) & 0.04(0.04) & -0.04(0.04)\\
        $\beta_\text{Suedalpin}$ & -0.57(0.02) & -0.02(0.04) & -0.11(0.04) \\
        $\beta_\text{Tertiaere Intrusiva und Extrusiva}$ & 0.26(0.13) & 0.49(0.12) & 0.43(0.13)\\
        $\beta_\text{Unterostalpine Decken}$ & -0.22(0.03) & 0.23(0.08) & 0.26(0.08) \\
        $\beta_\text{Unterpenninische Kristallindecken}$ & -0.41(0.03) & 0.01(0.04) & -0.03(0.05)\\
        $\beta_\text{Unterpenninische Sedimentdecken und -schuppen, Ophiolithe}$ & -0.51(0.02) & 0.01(0.05) &  -0.046(0.050)\\
    \bottomrule
    \end{tabular}
    }
    \label{tab:coeffs_tecto}
\end{table}        
        
\begin{table}[ht]
    \centering
    \caption{Fitted coefficients in the different modelling approaches for categories of land coverage.}
    \begin{tabular}{lllll}
    \toprule
        & lm inla & mixed & extended \\
        & mean(sd) & mean(sd) & mean(sd)\\
    \midrule       
        $\beta_\text{artificial}$ & reference & reference & reference \\
        $\beta_\text{grass vegetation}$ & 0.133(0.004) & 0.078(0.004) & 0.035(0.003) \\
        $\beta_\text{bush vegetation}$ & 0.028(0.008) & -0.030(0.006) &  -0.019(0.006)\\
        $\beta_\text{tree vegetation}$ & -0.039(0.004) & -0.072(0.004) & -0.038(0.004) \\
        $\beta_\text{without vegetation}$ & 0.006(0.010) & -0.032(0.009) &  -0.009(0.008)\\
        $\beta_\text{wetland}$ & -0.16(0.02) & -0.19(0.02) & -0.13(0.02) \\
    \bottomrule
    \end{tabular}
    \label{tab:coeffs_lc}
\end{table}

\begin{table}[ht]
    \centering
    \caption{Fitted Coefficients in the different modelling approaches for cumulative rainfall after Chernobyl. We rescaled the cumulative rainfall to meters to compute the models.}
    \begin{tabular}{lllll}
    \toprule
        & lm inla & mixed & extended \\
        & mean(sd) & mean(sd) & mean(sd)\\
    \midrule
        $\beta_\text{rainfall}$ & 6.5(0.2) & 6.5(1.6) & 5.1(1.8) \\
    \bottomrule
    \end{tabular}
    \label{tab:coeffs_rain}
\end{table}

\clearpage
\section{Maps of predictors}

\begin{figure}[h]
    \centering
    \includegraphics[width=0.8\linewidth]{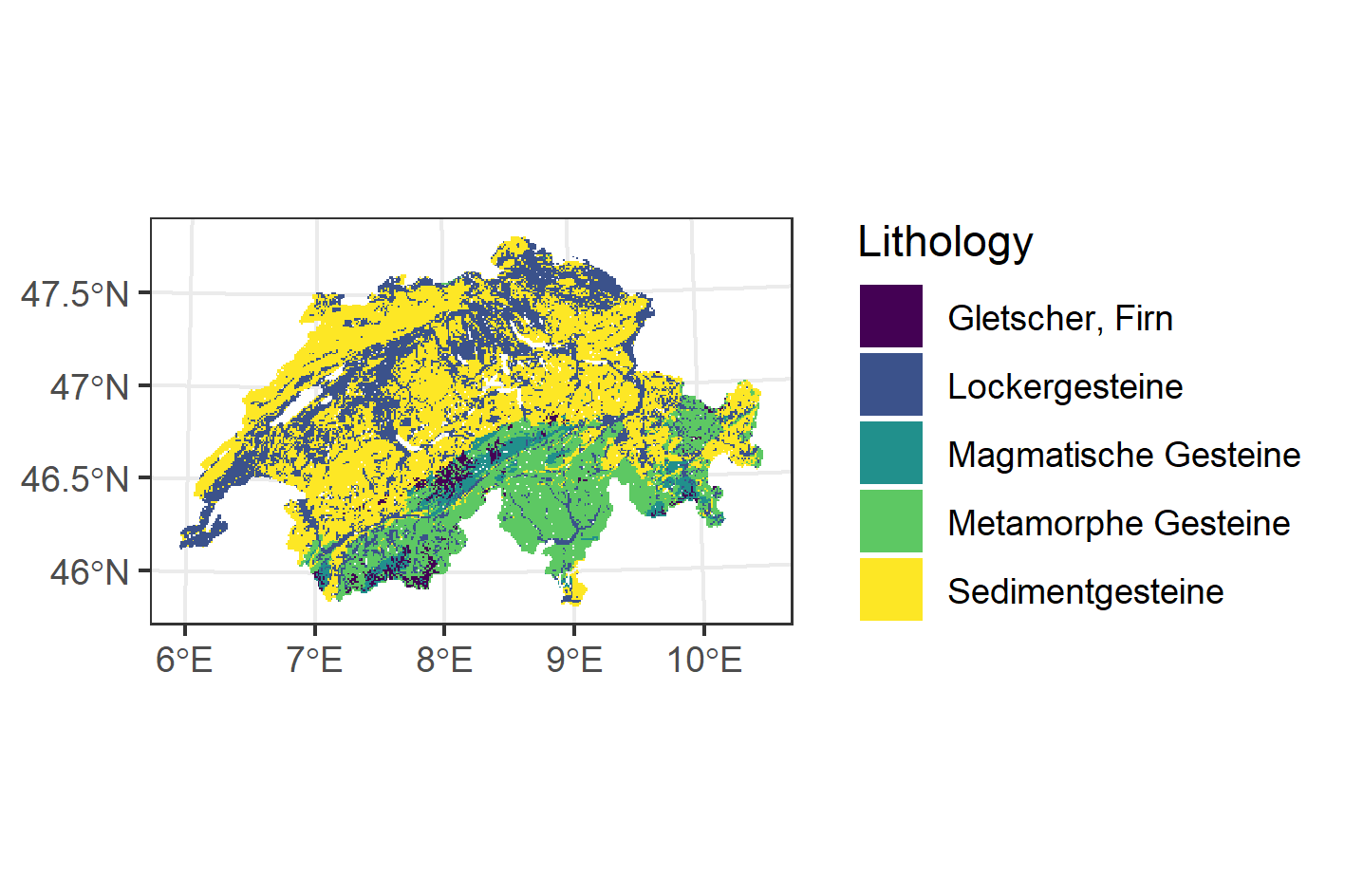}
    \caption{Map of Lithology. (Data source: swisstopo.)}
    \label{fig:litho}
\end{figure}

\begin{figure}[h]
    \centering
    \includegraphics[width=0.8\linewidth]{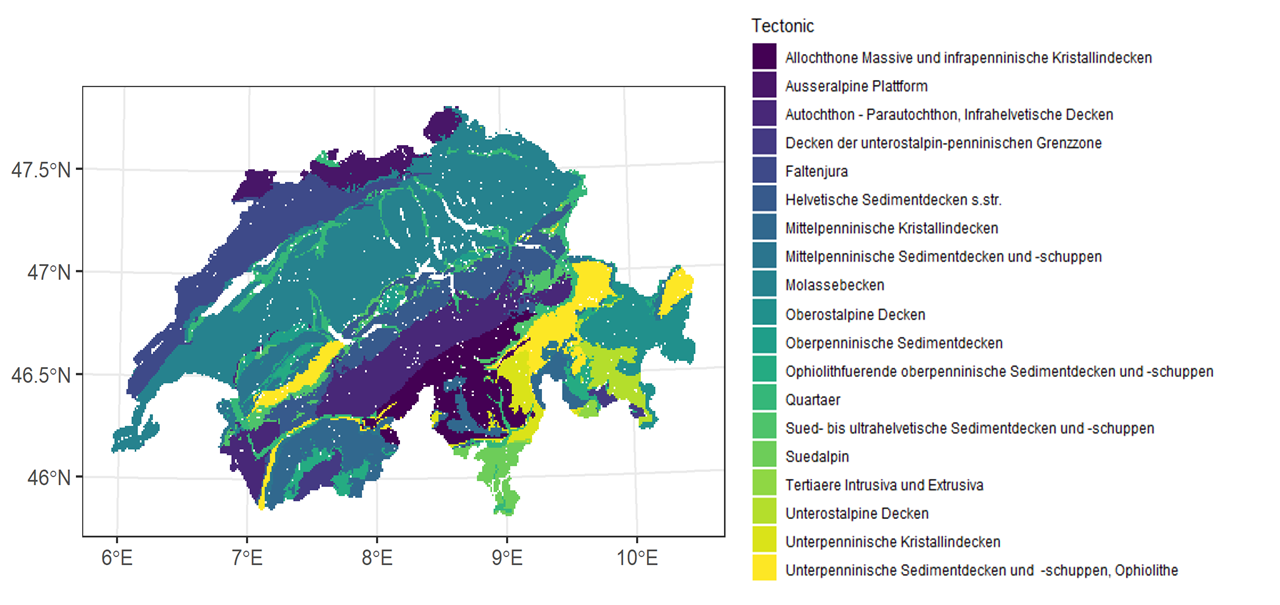}
    \caption{Map of Tectonic units.(Data source: swisstopo.)}
    \label{fig:tecto}
\end{figure}

\begin{figure}[h]
    \centering
    \includegraphics[width=0.8\linewidth]{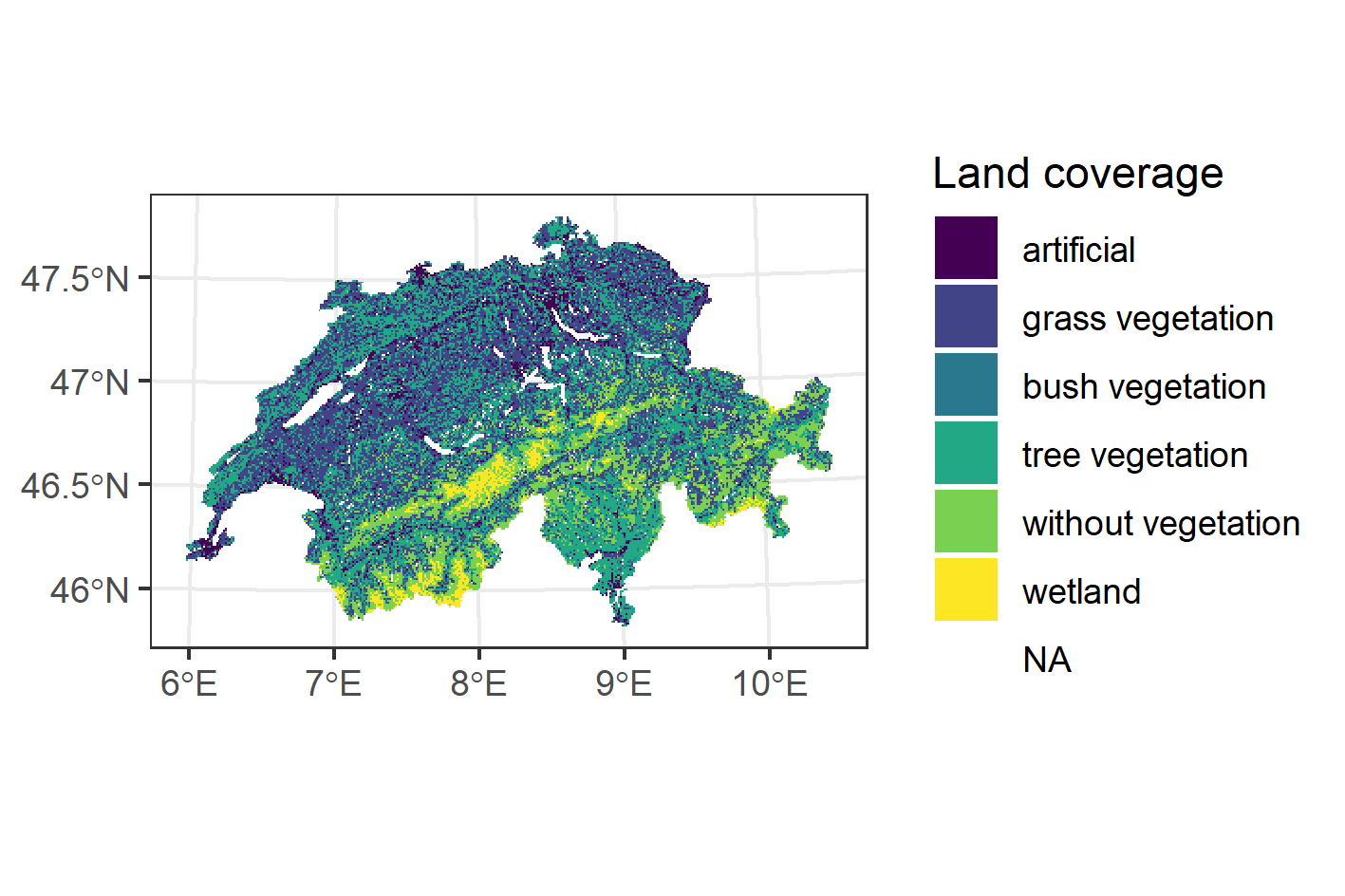}
    \caption{Map of Land coverage. (Data source: Federal Office for Statistics.)}
    \label{fig:landcover}
\end{figure}

\begin{figure}[h]
    \centering
    \includegraphics[width=0.8\linewidth]{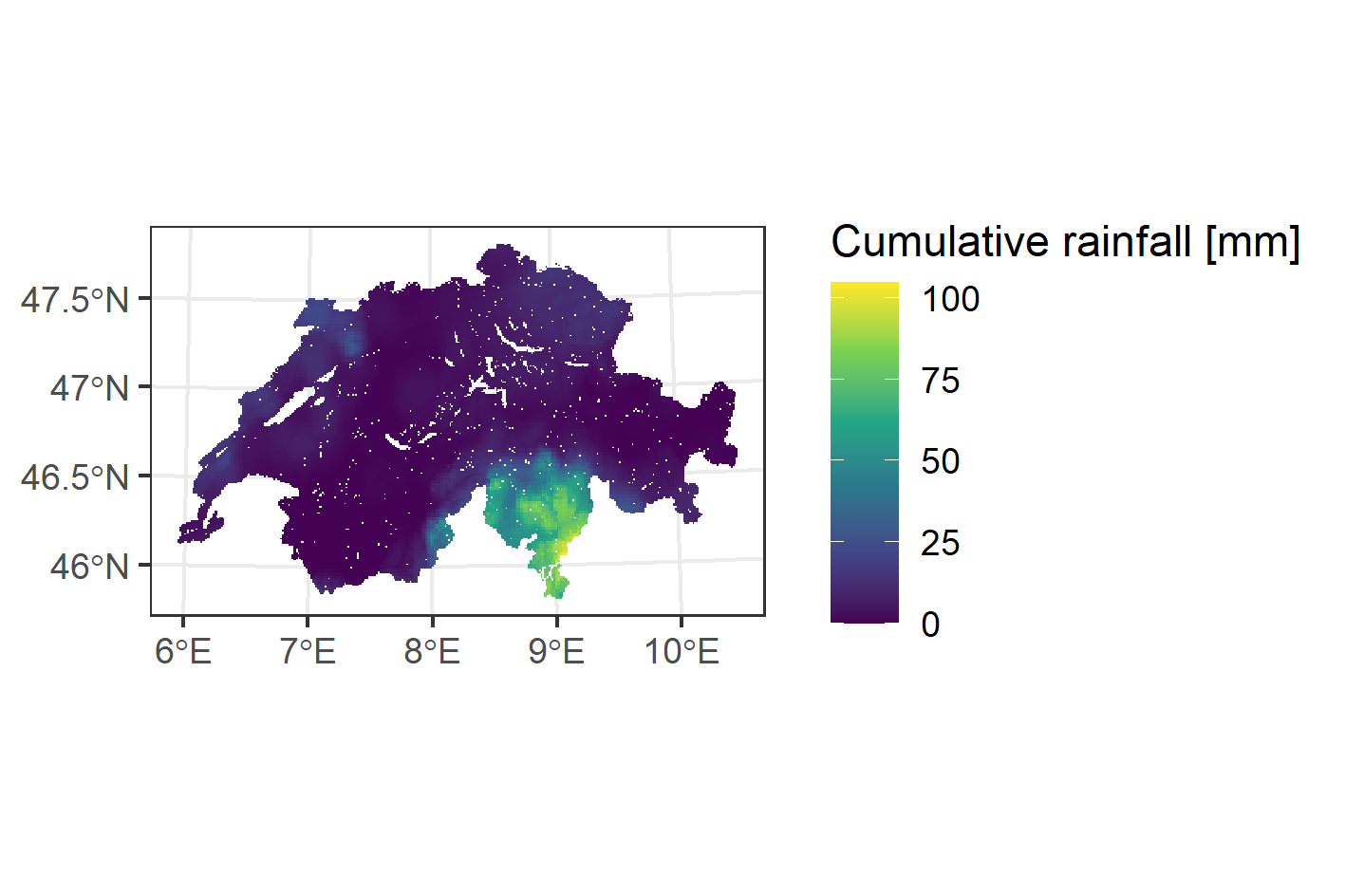}
    \caption{Map of cumulative rainfall from 30 April until 5 May 1986. (Data source: Meteosuisse)}
    \label{fig:rainfall}
\end{figure}

\end{document}